# Observation of dynamical degeneracy splitting for the non-Hermitian skin effect

Tuo Wan[1], Kai Zhang[2], Junkai Li[1], Zhesen Yang[3]* and Zhaoju Yang[1]*

[1]School of Physics, Interdisciplinary Center for Quantum Information, Zhejiang Province Key Laboratory of Quantum Technology and Device, Zhejiang University, Hangzhou 310027, Zhejiang Province, China

[2]Department of Physics, University of Michigan Ann Arbor, Ann Arbor, Michigan, 48105, United States

[3]Department of Physics, Xiamen University, Xiamen 361005, Fujian Province, China

*Email: yangzs@xmu.edu.cn; zhaojuyang@zju.edu.cn.

**Abstract**

The non-Hermitian skin effect is a distinctive phenomenon in non-Hermitian systems, which manifests as the anomalous localization of bulk states at the boundary. To understand the physical origin of the non-Hermitian skin effect, a bulk band characterization based on the dynamical degeneracy on an equal frequency contour is proposed, which reflects the strong anisotropy of the spectral function. In this paper, we report the experimental observation of both phenomena in a two-dimensional acoustic crystal, and reveal their remarkable correspondence by performing single-frequency excitation measurements. Our work not only provides a controllable experimental platform for studying the non-Hermitian physics, but also confirms the correspondence between the non-Hermitian skin effect and the dynamical degeneracy splitting, paving a new way to characterize the non-Hermitian skin effect.

In quantum physics, the restriction of the conservation of probability imposes that the Hamiltonian describing the physical system is always Hermitian. However, when the system interacts with external environments, the corresponding Hamiltonian can become non-Hermitian. In the past two decades, such non-Hermitian physics[1-4] has been extensively explored in condensed matter physics [5-7], photonics [8-26], phononics [27-38] and many other platforms [39-43]. Characterized by complex-valued eigenenergies and nonorthogonal eigenstates, non-Hermitian physics has offered many interesting features that cannot be found in Hermitian systems, such as exceptional points [4,10,13,19,22,23,29], non-Hermitian skin effect [24,31,35,41,43-51] and complex-energy braiding [52-54] with potential applications in sensing [55-58] and lasing [59,60].

Among these advances, the non-Hermitian skin effect (NHSE) has drawn a lot of attention. In one dimension, the NHSE imposes all the eigenstates onto the boundaries of the system, which can be well described by the non-Bloch band theory [44-51]. However, the situation in two dimensions is much more complicated. As depicted in Fig. 1, based on the localization dimension and the number of corresponding localized states, the two-dimensional (2D) skin modes can be classified into the skin corner mode [45] [Fig. 1(a)], the skin edge mode [61] [Fig. 1(b)], and the hybrid skin-topological corner mode [62] [Fig. 1(c)], with the number of relevant modes proportional to $L^2$, $L^2$ and $L$ ($L$ is the characteristic length of the polygon), respectively [61-63].

Combined with the fact that the number of the topological corner modes [64,65] [Fig. 1(d)] and topological edge modes [66] [Fig. 1(e)] that appear in different Hermitian topological systems are proportional to $L^0$ and $L^1$, one can generalize that the number of the localized states is the key point to distinguish the NHSE from the topological boundary states. For even higher dimensions, there is no generalized non-Hermitian band theory or classification of the skin modes that has been achieved yet.

Recently, a frequency criterion for the appearance of the NHSE has been proposed [67], i.e. the dynamical degeneracy splitting (DDS) on an equal frequency contour (EFC) which refers to the phenomenon that the EFC for a real-valued frequency has no common lifetime. Theoretically, the DDS can help us predict the presence or absence of the NHSE for that given frequency and explain which edge or corner contains skin modes. A typical application of the DDS is to explain the novel localization phenomenon in the geometry-dependent skin effect (GDSE) proposed recently in Ref. [61]. In the GDSE, the appearance or absence of the NHSE depends explicitly on the shape of the 2D geometry with the open boundary condition (OBC). For example, the NHSE may disappear under the square-geometry but appear in generic polygon geometry. Experimentally, the DDS can be detected in many spectral measurements, which provides an accessible method to verify the existence of the NHSE. Therefore, it is important to examine the presence of the DDS and its relationship to the NHSE (especially the GDSE) experimentally.

In this work, we experimentally demonstrate the GDSE in an acoustic lattice and unveil that the DDS plays the role of local-frequency criterion for the existence of the NHSE. First, we design a tight-binding model whose unit cell consists of four sites with detuned on-site potentials and non-Hermiticity. This model can be classified into the case of skin edge modes shown in Fig. 1(b) and support the GDSE, whose global eigenstates integrating over the entire spectrum condense on the edges with a specific direction. We next map this non-Hermitian theoretical model into an acoustic lattice and perform the local acoustic measurements. Through summing over the operating spectrum, we observe the non-Hermitian skin edge modes in the triangular lattice but not in the square lattice. The spectral function of the EFC as a function of the operating frequency is detected to investigate the presence or absence of the DDS. Our measured result shows that when the EFC has (no) DDS, the corresponding eigenstates under the OBC at that frequency exhibit (no) GDSE, which verifies the deep connection between the DDS and GDSE.

We start from a 2D tight-binding model with four sites (A, B, C, D) in a unit cell, as shown in Fig. 2(a). The color of blue (red) indicates the on-site potential of $m$ $(-m)$. The thin (thick) circle indicates the on-site loss with the strength of 0 $(\gamma)$. Therefore, the Hamiltonian of this model can be written as:

$$H = \sum_{<n,m>} t c_n^\dagger c_m + \sum_n (M_n + \Gamma_n i) c_n^\dagger c_n \tag{1}$$

where $c_n^\dagger$ ($c_n$) is the creation (annihilation) operator, $t$ is the nearest-neighbor coupling strength, the on-site potential $M_n = -m(+m)$ for sites $n \in \{A, D\}(\{B, C\})$, the on-site loss $\Gamma_n = 0(-\gamma)$ for sites $n \in \{A, B\}(\{C, D\})$. In the following, we set $m = 2t$ and $\gamma = 2t$ for further studies.

One evidence for the existence of the GDSE lies within the spectra under the periodic boundary condition (PBC) and OBC in the complex-energy plane[61]. We calculate the complex-energy spectra for two finite lattices in square and triangular shapes, as depicted in Fig. 2(b, c). The blue region indicates the energy spectrum under PBC and the red dots represent the one under OBC. We can find that (i) the PBC spectrum covers a non-zero spectral area in the complex-energy plane, which suggests the existence of the NHSE [61]; (ii) the distribution of the OBC spectrum depends explicitly on the open-boundary geometries, which indicates that the NHSE is geometry-dependent. Another evidence for the presence of the GDSE is to directly investigate the averaged field distribution over all the eigenstates, defined as $W(n) = 1/N \sum_m |\psi_m|^2$, where $n, m$ and $N$ indicate the site index, eigenstate number and the total number of sites, respectively. It can be seen from Fig. 2(e) that, for the triangle shape, the field distribution shows higher intensity in the oblique edge, which displays the existence of NHSE; When it turns to the square shape, as shown in Fig. 2(d), the field pattern becomes uniformly distributed, indicating the absence of the NHSE under this shape. Note that in both two cases, there is no localized state at the horizontal and vertical edges.

To experimentally verify the geometry-dependent NHSE that exists in our proposed model, we map the non-Hermitian Hamiltonian into acoustic systems. The unit cell consists of four sonic resonators, whose eigenfrequency can be artificially tuned by changing the length of the resonator and non-Hermiticity can be introduced by inserting sound-absorbing material into the resonators [32]. The nearest-neighboring resonators couple to each other by smaller sonic waveguides. The schematic is shown in the inset in Fig. 3(a). The first evidence for the existence of the GDSE comes from the direct simulation of the spectra of the acoustic systems. As shown in Fig. S2(a, b) in the Supplementary Materials [68], section 2, the spectra show non-zero spectral areas for both square and triangular geometries and are clearly geometry-dependent, suggesting the existence of GDSE in the acoustic system. The second evidence lies within the integrated eigenstates from simulations for square and triangular samples. As can be clearly seen in Fig. S2(d), there exist skin edge modes at the oblique edge for the finite triangular sample, whereas for the finite square sample in Fig. S2(c), the field intensity is uniformly distributed over the lattice.

Through measuring the local density of states [69,70] in the acoustic system, we can experimentally retrieve the field distributions with respect to the frequency. Then by simply summing the field distributions over the operating frequency ranging from 4650Hz to 5300Hz, we can visualize the integrated field patterns from all the supported eigenstates. The result is presented in Fig. 3(c, d). We find that our acoustic system based on the non-Hermitian Hamiltonian Eq. 1 supports the GDSE. As shown in panel (c), the skin edge modes occur at the oblique edge of the triangular sample, whereas the integrated field pattern is uniformly distributed in the square lattice [panel (d)]. This central observation of the skin edge modes agrees well with the simulation result shown in Supplementary Materials, Fig. S2(c, d). The averaged intensities in the bulk and at the edges for different samples with the frequency ranging from 4650Hz to 5300Hz are shown in panel (e, f). The higher intensity at the oblique edge in the triangular sample and the uniformly distributed intensity in the square sample together verify the existence of the GDSE. Note that the high-intensity peak centered at 5100Hz corresponds to the right non-zero spectral area (2< $E$ <2.82), which indicates the existence of skin edge modes. In contrast, the measured intensity at the left non-zero spectral area (-2.82<E<-2, ~4800-4950Hz) is much weaker than the high-intensity peak at the right spectral area (2< $E$ <2.82) and equivalent to that in the bulk [blue shading in Fig. 3(e)]. The reason is the skin edge modes within this left spectral area have eigenenergies with larger imaginary parts, which result in the fast decay of the excited waves and small measured intensity.

To gain more insight into the reason why NHSE happens, we next discuss the DDS for a given frequency. For a given real-valued frequency, the density of states (DOS) in the Brillouin zone (BZ) can be obtained by the spectral function:

$$A(E, \mathbf{k}) = -\text{Im } Tr[1/(E + i\eta - H(\mathbf{k})] \qquad (2)$$

where $Tr$ represents the trace of a matrix, $\eta$ is a small constant, $E_1 = 2.7$ is a reference real energy, $H(\mathbf{k})$ is the Hamiltonian of Eq. (1) in the momentum space. We calculate $A(E, \mathbf{k})$ of our model for different values of $E$ [indicated by dashed lines in Fig. 2(b, c)], and investigate the relationship with the GDSE. We present the result of $A(E, \mathbf{k})$ in Fig. 4(a) with the energy of $E_1 = 2.7$. The colorbar represents the DOS and the grey dashed line represents the EFC. It can be seen that the DOS is mirror symmetric about the $k_x$ and $k_y$ axis. However, Fig. 4(a) exhibits unbalanced DOS distribution on the grey dashed line, showing the uneven broadening of EFC, which indicates the presence of DDS. In contrast, there is no DDS in Fig. S4(a, c, i) with the energy of $E_2 = -3.3, -1.7$ and $4.5$. More results of the spectral functions can be found in Supplementary Materials, section 5.

Now we use the DDS on the EFC to explain whether an OBC geometry exhibits the NHSE. We consider a launched wave packet with the center momentum $\mathbf{k}_i = (k_{x0}, k_{y0}) = (0.86, -\pi)$, where the DOS is at a high level indicating a longer lifetime. In general, when this wave packet strikes a spatial boundary, the center momentum may be scattered to a new momentum with the same energy. However, things get different when there is no scattering channel [67]. In our model, when this wave packet strikes the vertical edge, the momentum of the wave packet is scattered from $(k_{x0}, k_{y0})$ to $(-k_{x0}, k_{y0})$, and two momentum points share the same DOS, namely, the same lifetime. Therefore, the ordinary scattering process can happen. However, when the edge is oblique, the wave packet with the momentum of $\mathbf{k}_i$ will be scattered to the one with the momentum of $\mathbf{k}_s$, whose DOS in EFC is far smaller than that in $\mathbf{k}_i = (k_{x0}, k_{y0})$, so the wave packet can not be scattered, and it has no choice left but localizes at the edge, and consequently forms the skin edge mode.

In experiments, by measuring the real-space acoustic pressure distribution and Fourier transforming it to the momentum space, we can retrieve the spectral function $A(E, \mathbf{k})$ and observe the presence or absence of the uneven broadening (or DDS) of the EFC. The result with the frequency of 5140Hz is presented in Fig. 4(b) and agrees well with the numerical result with the energy of $E_1 = 2.7$ shown in panel (a). More results of the measured spectral functions can be found in Fig. S4. The experimental results are consistent with the theoretical results, confirming the existence of DDS on the EFC.

To verify the correspondence between the DDS on the EFC and the NHSE, we further measure the field distribution at the different frequencies. The operating frequencies of 5140Hz and 4776Hz correspond to the eigenenergies of $E_1 = 2.7$ and $E_2 = -3.3$, which lie in and out of the non-zero spectral area indicating the presence and absence of the NHSE. We observe that at the frequency of 5140Hz, the field distribution localizes at the oblique edge of the triangular geometry, as shown in Fig. 4(e), while no localized mode is found at the edges of the square geometry, as shown in Fig. 4(c). In contrast, at the frequency of 4776Hz, the DOS on the EFC [shown in Fig. S4(b)] is much smaller than that at the frequency of 5140Hz. We observe no skin edge modes in Fig. 4(d, f) and the field distribution in the triangular sample shows no strong localization effect. These key observations demonstrate the presence and absence of the GDSE with and without the DDS, which unveils the significant correspondence between the two unique non-Hermitian phenomena.

In summary, we have studied a new type of 2D NHSE, the so-called GDSE [61], and experimentally demonstrated the existence of the DDS on the EFC, which can not only explain the physical origin of the NHSE but also predict which edges can host the skin edge modes. Since the DDS characterizes the bulk property of the non-Hermitian system, and the NHSE manifests at the boundary, they can be regarded as a novel type of bulk-boundary correspondence in non-Hermitian systems. Based on this correspondence, a new phenomenon called directional invisibility has been proposed recently [67], which suggests a possible direction for further experimental study of the higher dimensional NHSE.

**Note:** During our preparation of this work, recently we notice two preprints reporting the GDSE[71,72] in the acoustic and mechanical systems without manifesting its relationship with the DDS.

**Acknowledgements**
This research is supported by the National Key R&D Program of China (Grant No. 2022YFA1404203), National Natural Science Foundation of China (Grant No. 12174339), Zhejiang Provincial Natural Science Foundation of China (Grant No. LR23A040003) and Excellent Youth Science Foundation Project (Overseas). Z.Y. was also supported by the National Science Foundation of China (Grant No. NSFC-12104450), the fellowship of China National Postdoctoral Program for Innovative Talents (Grant No. BX2021300) and the fellowship of China Postdoctoral Science Foundation (Grant No. 2022M713108).

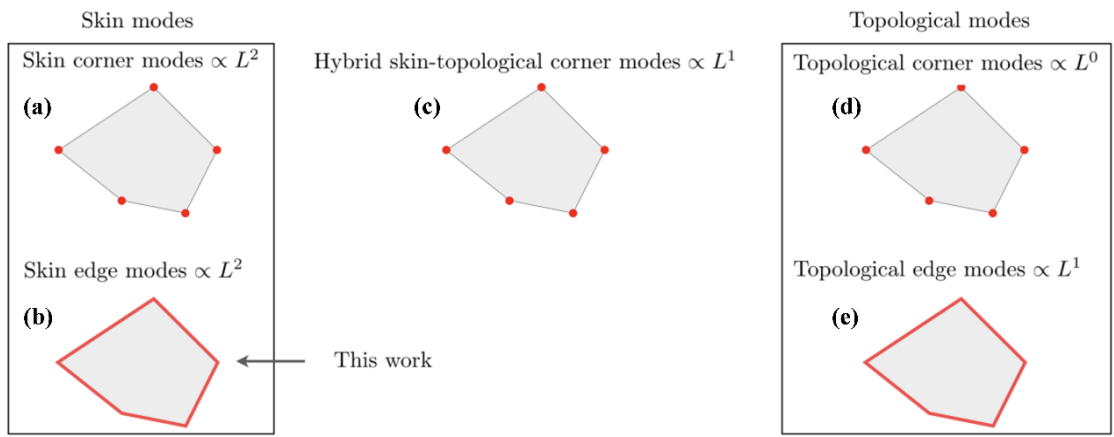

Figure 1: The illustration of 2D non-Hermitian skin modes and topological edge modes. The red dots or lines indicate that the modes are localized at corners or edges, and $L$ represents the characteristic length of the polygon.

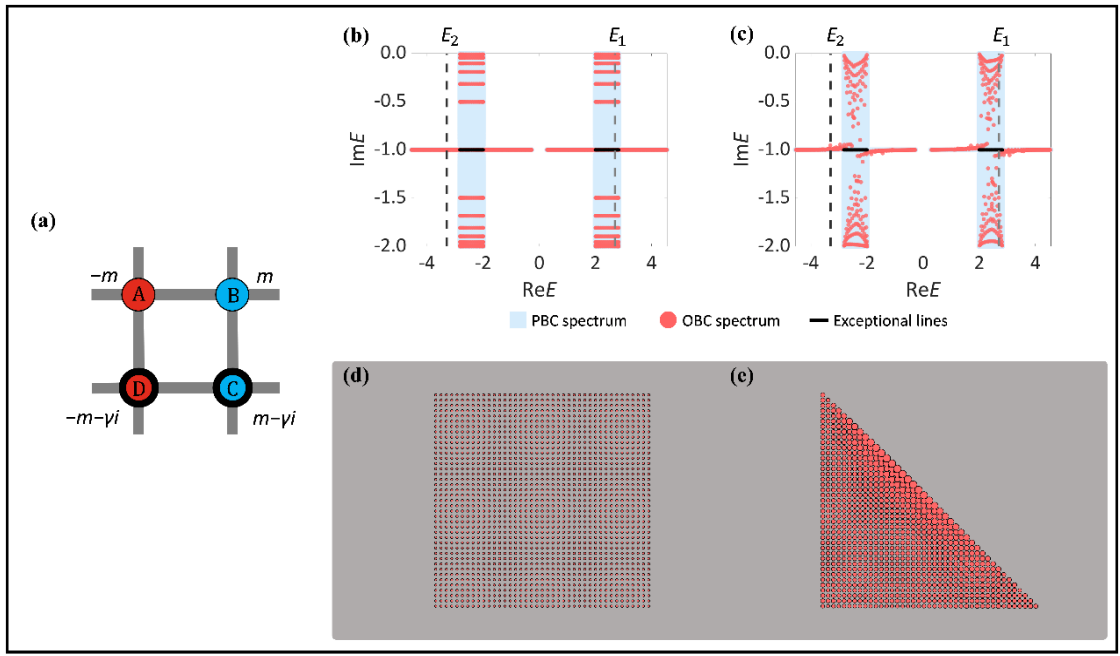

Figure 2: Non-Hermitian tight-binding model. (a) Schematic of the unit cell. The color of blue (red) indicates the on-site potential of $m$ ($-m$). The thin (thick) circle indicates the on-site loss with the strength of $0$ ($\gamma$). (b, c) Eigenenergies on the complex-energy plane for the square and triangular geometries, where the blue shadings represent the PBC spectrum and the red dots represent the OBC spectrum. (d-e) Numerical result of the summed eigenstates $W(n)$ for the square (d) and triangular (e) geometries. The size of the circle denotes the strength of $W(n)$. The other parameters are set to be $m = 2t, \gamma = 2t$.

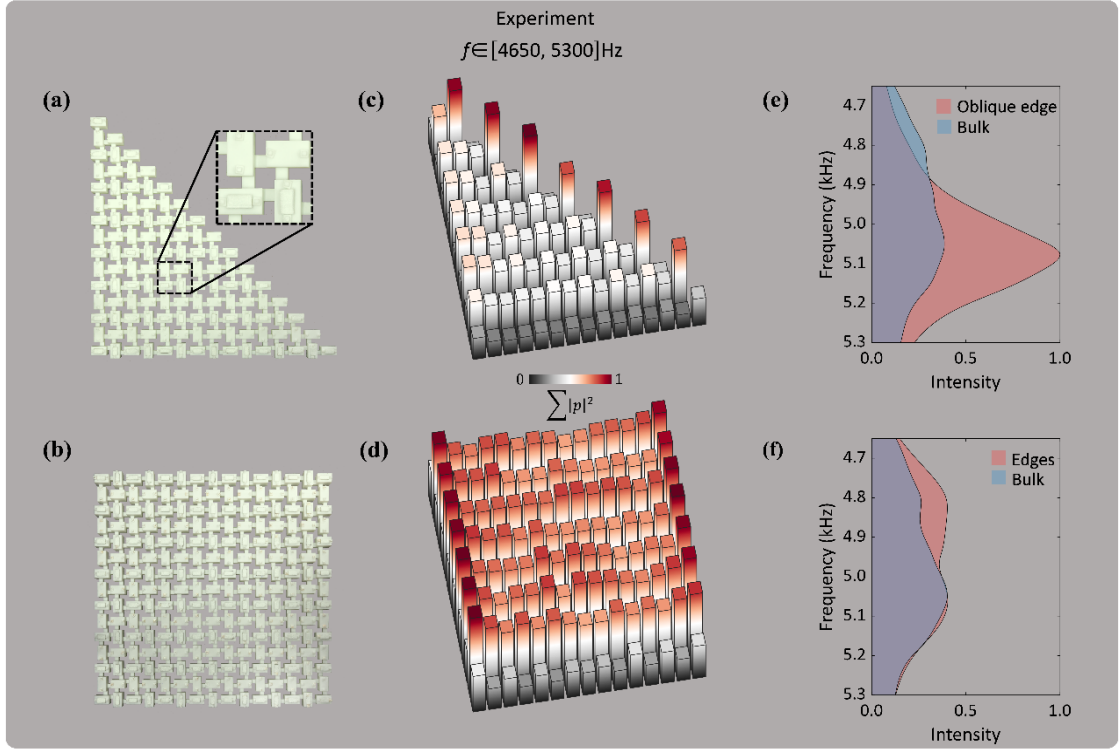

Figure 3: Observation of the GDSE. (a, b) Photographs of the triangular and square samples. The inset in panel (a) shows the unit cell. (c, d) Experimental results of the summed acoustic intensity distributions $\sum|p|^2$ for the triangular and square geometries. The colorbar indicates the strength of $\sum|p|^2$. (e) Measured intensity spectrum on the oblique edge (red shading) and in the bulk (blue shading) of the triangular sample. (f), Measured intensity spectrum on the edges (red shading) and in the bulk (blue shading) of the square sample. Intensities in (e, f) are normalized by the maximal value of the oblique edge's intensity. The measured high intensity at the oblique edge of the triangular indicates the existence of the NHSE.

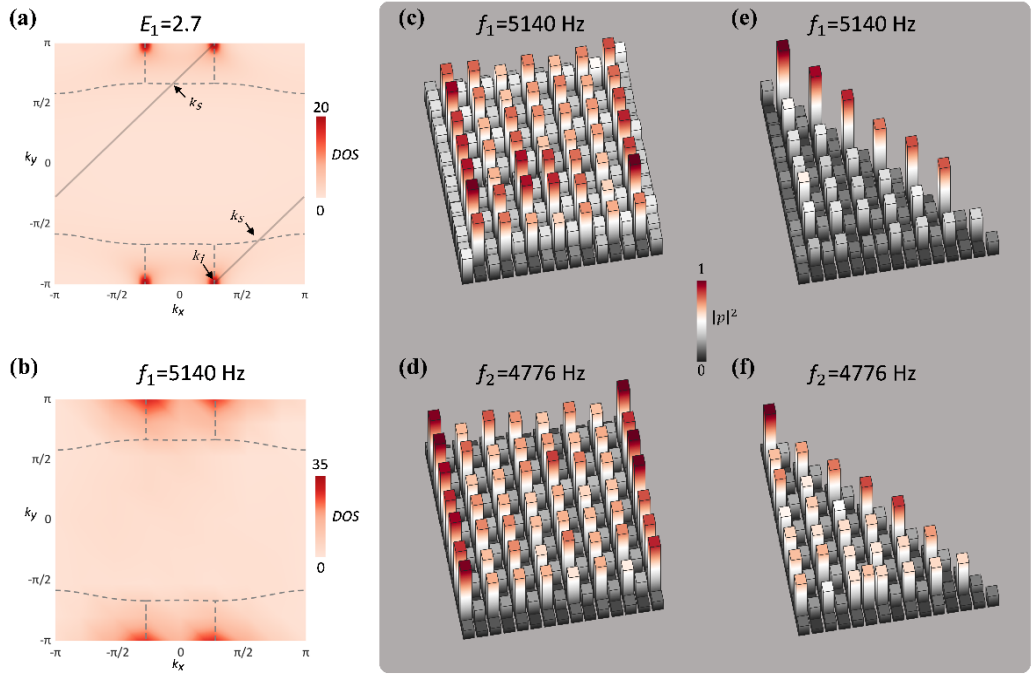

Figure 4: DDS on the EFC and GDSE. (a) Calculated spectral function $A(E, \mathbf{k})$ at the energy of $E_1 = 2.7$. Grey dashed lines denote the EFC. The incident wave with the momentum of $k_i$ can be back-scattered to the wave with the momentum of $k_s$ when striking the oblique edge. (b) Measured spectral function at the frequency of $f_1 = 5140$ Hz corresponding to panel (a). (c, d) Measured acoustic intensity distributions for the square sample at the frequency of $f_1 = 5140$ Hz and $f_2 = 4776$ Hz, which correspond to the energy of $E_1 = 2.7$ and $E_2 = -3.3$ respectively. (e, f) Measured acoustic intensity distributions for the triangular sample at the frequency of $f_1 = 5140$ Hz and $f_2 = 4776$ Hz, which correspond to the energy of $E_1 = 2.7$ or $E_2 = -3.3$.

# Supplementary Materials

Tuo Wan[1], Kai Zhang[2], Junkai Li[1], Zhesen Yang[3*] and Zhaoju Yang[1*]

[1]School of Physics, Interdisciplinary Center for Quantum Information, Zhejiang Province Key Laboratory of Quantum Technology and Device, Zhejiang University, Hangzhou 310027, Zhejiang Province, China

[2]Department of Physics, University of Michigan Ann Arbor, Ann Arbor, Michigan, 48105, United States

[3]Department of Physics, Xiamen University, Xiamen 361005, Fujian Province, China

*Email: yangzs@xmu.edu.cn; zhaojuyang@zju.edu.cn.

## Section 1. Experimental Set-up

The experimental sample is shown in Fig.S1. The resonators with positive (negative) potentials are in the size of $3.4(3.5)\text{cm} \times 1.9\text{cm} \times 1.9\text{cm}$. The loss is induced by inserting sound-absorbing material in the $1.14\text{cm} \times 2.1\text{cm}$ holes on the top of the resonators (see the inset in Fig. S1, red rectangles). There are two pinholes in each resonator for placing the microphone (square holes, $3\text{mm} \times 4\text{mm}$) and detector (circle holes, r=1.5mm). They are covered by plugs when they are not being used. The nearest-neighboring resonators are connected by additional waveguides with the size of $0.7\text{cm} \times 0.7\text{cm} \times 1.5\text{cm}$.

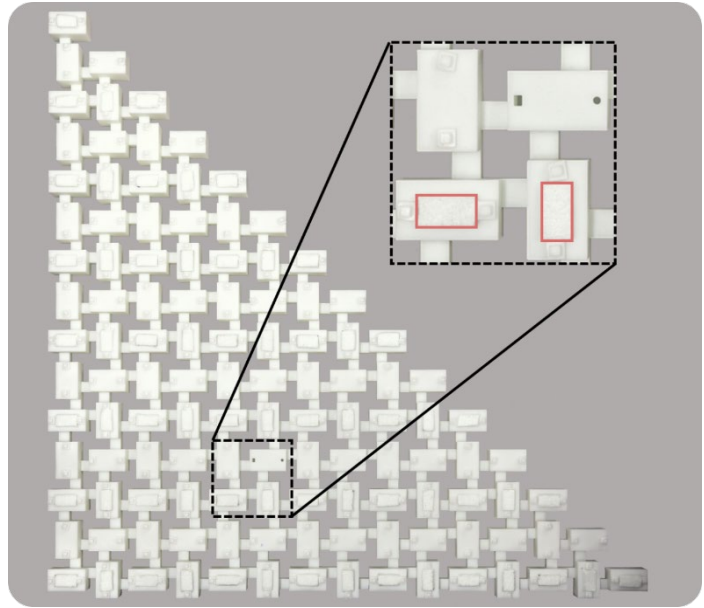

Figure S1. Triangular sample. The inset shows a unit cell. The red rectangles enclose the sound-absorbing materials.

During our measurement of LDOS, we insert a speaker (Knowles SWFK-31736-000 with a diameter of 1.4mm) into rectangle pinholes as an acoustic source and place the microphone (Brüel Kjær 4187) into the circular hole of the same resonator as a detection probe. The detected signals of every site are transferred to a computer through a data analyzer (Brüel Kjær 3160-A-042).

To obtain the spectral intensity in the momentum space, we first measure real-space acoustic pressure distributions and then perform direct Fourier transform operation. In this measurement, a reference microphone (Brüel Kjær 2670) is used to retrieve the phase information of the states.

## Section 2. Simulation results for our acoustic model

Here, we perform simulations of our acoustic model in comparison to the tight-binding model. The loss is introduced by adding an imaginary part to the sound velocity $c = c_0(1 + i\gamma)$, where we take $c_0 = 340\text{m/s}$, $\gamma = 0.05$ and sound density $\rho = 1.29\text{kg/m}^3$. The value of $\gamma$ is determined by comparing the eigenvalue of a resonator with the scattering boundary condition (1.14cm × 2.1cm) on the top of a reaonator and the eigenvalue of a resonator filled with the complex-velocity air.

We calculated the eigen-frequencies and $W(n) = 1/N \sum_m |\psi_m|^2$ of the acoustic crystals in the square and triangular samples and present them in Fig.S2. The spectra with non-zero spectral area [Fig. S2(a, b)] and the intensity distributions localized at the oblique edge [Fig. S2(c)] indicate the existence of the GDSE. Note that the frequency shift in our acoustic measurement can be compensated by tuning the inserting length of the absorbing material. More details can be found in our earlier work [1].

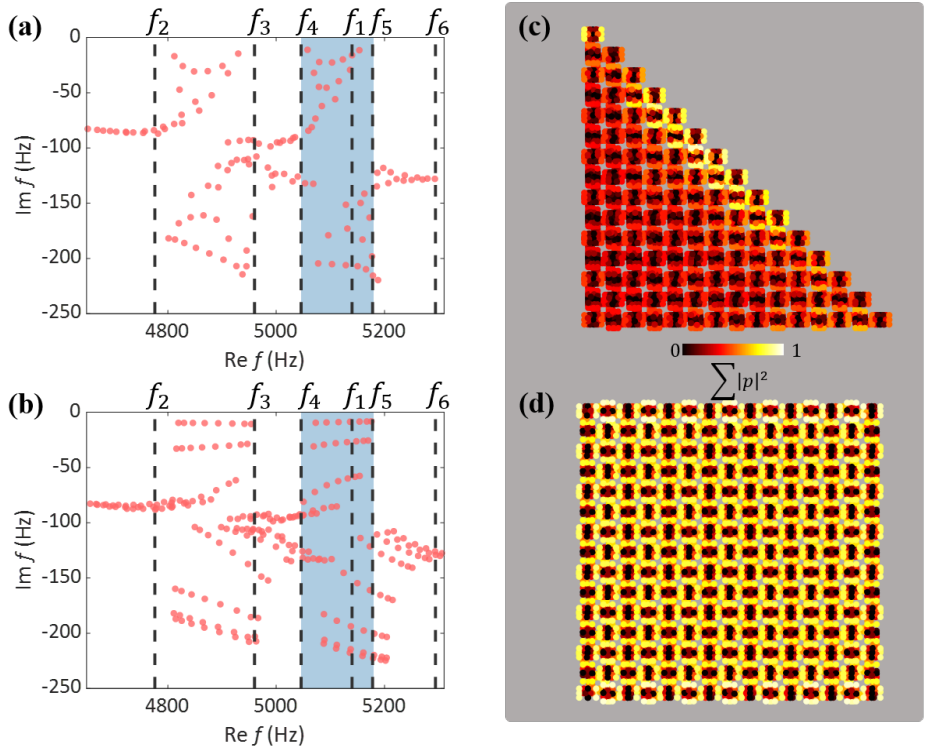

Figure S2. Simulation results. (a, b) The simulated spectra for the triangular and square samples. Blue shadings mark the frequencies of upper bands. There exists the NHSE in trigular sample. The labeled $f_4$ is the leftmost boundary of the blue shading. The $f_{2-6}$ are the operating frequencies that will be discussed in section 5. (c, d) The simulated summed eigen intensity fields $W(n)$ for the square and triangular samples.

**Section 3. The band structure in the Brillouin zone**

In this section, we present the band structure of the Hamiltonian Eq. 1 in the main text. The band structure exhibits exceptional lines (black lines).

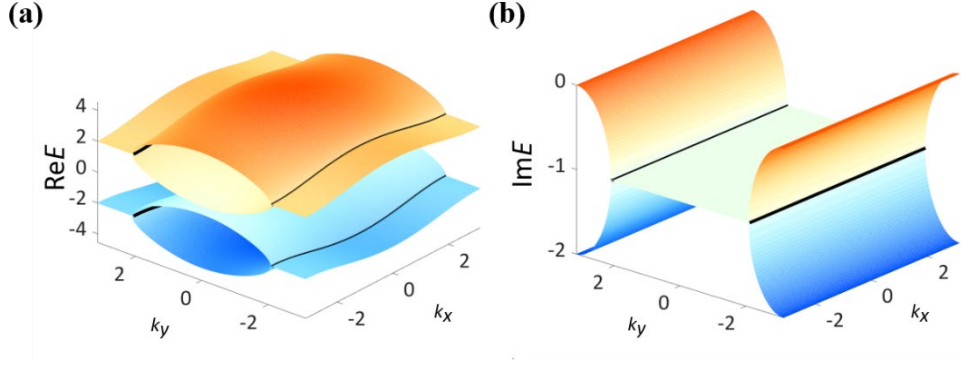

Figure S3: The band structure of the Hamiltonian Eqn. 1 in the main text. (a) real part of the band structure, where black curves denote the exceptional lines. (b) Imaginary part of the band structure, where the black curves denote the exceptional lines.

**Section 4. More results for spectral functions**

In this section, we provided more results of the spectral function $A(E, \boldsymbol{k})$ for different energies. The results are shown in Fig. S4. The energies or frequencies are makred in black dashed lines in Fig. S2(a, b). Compared with the cases supporting the DDS [see Fig. S4(e, g) and Fig. S4(f, h) for theoretical and experimental results], the spectral functions in the energies without the DDS [Fig. S4(a, c, i) and Fig. S4(b, d, j)] show much smaller spectral intensity than those in Fig. S4(c, g) and Fig. S4(f, h). The measured results in the lower panels agree well with the theoretical results.

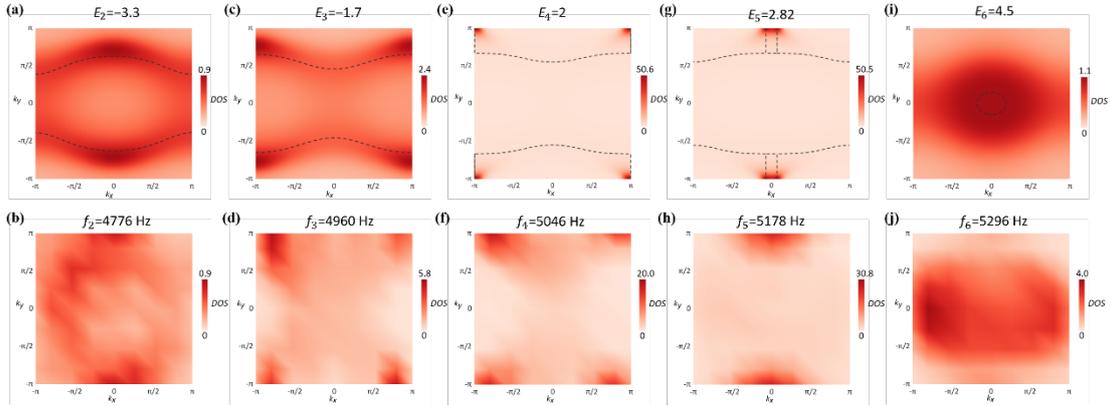

Figure S4: Theoretical and experimental results of the DDS. (a, c, e, g, f), Theoretical results of the spectral function $A(E, \boldsymbol{k})$. The EFC is marked in dashed lines. (b, d, f, h, j) Experimental results of the spectral functions with different frequencies corresponding to the upper panels (a, c, e, g, f). The operating frequencies are also marked in Fig .S2(a, b).